\documentstyle[twocolumn,eqsecnum,aps]{revtex}
\begin{document}
\draft
\title{
\begin{flushright} {\normalsize NT@UW-98-7}\\
\end{flushright}
Mesons in Nuclei in the Light-Front Mean Field Approximation and
Deep-Inelastic Scattering}
\author{M. Burkardt}
\address{Department of Physics\\
New Mexico State University\\
Las Cruces, NM 88003-0001\\U.S.A.}

\author{G. A. Miller}
\address{Department of Physics, Box 351560\\
University of Washington \\
Seattle, WA 98195-1560\\U.S.A.}
\date{\today}
\maketitle
\begin{abstract}
A light-front treatment for the scalar and vector meson 
momentum distribution functions is developed using a model in which the nucleus
is treated  a static source of radius $R$. The limit $R\to \infty$ corresponds
to infinite nuclear matter. In this limit, 
the scalar mesons are shown
to carry a vanishing momentum fraction of the nuclear plus momentum
$k^+$, but the vector mesons 
carry a significant fraction, all occuring at the experimentally
inaccessible value of $k^+=0$.
This confirms   earlier work. A study of the $R$ dependence for sizes
of real nuclei shows that a fraction of the scalar and vector mesons
could be observable.
\end{abstract}
\narrowtext
\section{Introduction}
It has been known that
there is a significant difference between the parton distributions
of free nucleons and nucleons in a nucleus, since the famous EMC experiment
\cite{emc}.
The EMC-effect can be explained if the momentum distribution of
valence quarks is slightly shifted towards smaller values of $x$
for nucleons in a nucleus. One way to achieve this result is
by having mesons carry a larger fraction of the momentum in the 
nucleus\cite{emcrevs}. While such a model explains the shift in the valence
distribution, one obtains at the same time a meson
(i.e. anti-quark) distribution in the nucleus,
which is strongly enhanced compared to free nucleons and which should be
observable  in Drell-Yan experiments\cite{dyth}.
However, no such enhancement has been observed experimentally\cite{dyexp}.
The significance of this absence has been emphasized in Ref.~\cite{missing}.
 
The so-called EMC effect is a subtle one. For example, its size is typically 
about 10\% which is of the same order as relativistic effects.  
Motivated by the need to include a relativistic treatment of the motion of the
nucleons, which is also  consistent with the information derived using
conventional nuclear dynamics, one of us attempted to construct a light 
front treatment of nuclear physics\cite{jerry}.
The light front is relevant because
structure functions depend on the
Bjorken variable which is a ratio of the plus-momentum $k^0+k^3$
of a quark to that 
of the target. The calculations of 
Ref.~\cite{jerry}, using a Lagrangian in which Dirac nucleons
are coupled to massive scalar and vector mesons\cite{bsjdw},
treated the example of infinite 
nuclear matter within the mean field approximation. In this case, 
the meson fields are constants in both space and time. This means that the 
momentum distribution has support only at $k^+=0$. Such a distribution would
not be accessible experimentally, so that the suppression of the plus-momentum
of valence quarks would not imply the existence of a corresponding testable
enhancement of anti-quarks. However, 
it is necessary to ask if the result is only a artifact of the infinite nuclear
size and of the mean field approximation.

In this note, we investigate the dependence on nuclear size by 
explicitly constructing  the scalar and vector meson states for
a model in which the nucleus is represented as a static source of radius
$R$ and 
mass $M_A$. Since $M_A$ is very large
for large $A$, the nucleus acts as an extended static 
source. It should be emphasized that we do not
assume that individual nucleons are infinitely 
heavy to obtain this result. In the mean field approximation, the nucleus 
acts as a static source if it is infinitely heavy. Then a coherent state is 
the ground state of the
Hamiltonian.
 The organization of the paper is as follows. The model for the treatment
of scalar mesons is defined using the rest-frame (equal time formulation) in
Sect.~II. The light front LF treatment of that system is made in Sect.~III, in 
which the plus-momentum distribution function is computed. We study 
vector mesons 
in the LF treatment in Sect.~IV, and the distribution functions are
 obtained. The separate treatment of scalar and vector meson effects 
is made to simplify the discussion. Sect.~V contains a summary of the 
results and their 
possible implications. The derivation of a term in the light front Hamiltonian
that includes  the recoil momentum of the 
heavy source is made in an Appendix.

\section{Scalar  Meson Distribution in the Rest-Frame Formulation}
Before we proceed to the LF formulation, let us first analyse
the scalar  meson distribution in the mean field approximation in
a rest frame. This  simple  
calculation is of help in understanding the LF calculation.


Our 
model here is that of scalar mesons, coupled to a large static 
nucleus represented by a scalar  source $J({\vec r})$.
Such a system is described by the Lagrangian density
\begin{equation}
{\cal L}= \frac{1}{2}\partial_\mu \phi \partial^\mu \phi
-\frac{m^2_S}{2}\phi^2 + J \phi .
\label{eq:lager}
\end{equation} 
Canonical quantization proceeds by using 
\begin{equation}\phi(\vec{r})=\int {d^3k \over(2\pi)^{3/2}
\sqrt{2\omega_{\vec{k}}}}
\left[a_{\vec{k}}e^{i\vec{k}\cdot\vec{r}}+a_{\vec{k}}^\dagger
e^{-i\vec{k}\cdot\vec{r}}\right],
\end{equation}
where $\omega_{\vec k} = \sqrt{m_S^2+{\vec k}^2}$.
Then the  Hamiltonian reads
\begin{equation}
H = \sum_{\vec k} \left[
\omega_{\vec k} a^\dagger_{\vec k} a_{\vec k}
-\frac{1}{\sqrt{2\omega_{\vec k}}}
\left( \tilde{J}^*({\vec k})a^\dagger_{\vec k}
+ \tilde{J}({\vec k})a_{\vec k}\right)\right]
\label{eq:h0}
\end{equation}
where 
\begin{equation}
\tilde{J}({\vec k}) =\tilde{J}^*({-\vec k})
= (2\pi)^{-3/2}\int d^3r e^{-i{\vec k}\cdot{\vec r}}J({\vec r}).\label{jk}
\end{equation}
As a specific model, let us consider a spherical nucleus
of radius $R$ with constant density, i.e.
\begin{equation}
J({\vec r}) = J_0 \Theta (R-|{\vec r}|).
\label{eq:j}
\end{equation}
The quantity $J_0$ can be thought of as arising from the product
of a scalar meson coupling constant $g_S$ and an appropriate scalar density
$\rho_S$ so that 
\begin{equation}
J_0=g_S\rho_S.
\end{equation}

The ground state of $H$ (\ref{eq:h0})
can be obtained by means of a coherent state ansatz, with the result 
\begin{equation}
|\psi_0 \rangle \propto \left[ \prod_{\vec k}
\exp \left( \frac{\tilde{J}^*({\vec k})a^\dagger_{\vec k}
}{\sqrt{2\omega_{\vec k}^3}}\right)
\right] |0\rangle. \label{gsa}
\end{equation}
This state is an eigenfunction of $H$ with eigenvalue 
(binding energy)
\begin{equation}
E_0 = -\frac{1}{2}\int d^3k \frac{ \mid\tilde{J}({\vec k})\mid^2}
{\omega_{\vec k}^2}.
\label{eq:e0}
\end{equation}
This can be re-expressed in terms of an integral over coordinate space:
\begin{equation}
E_0 = -\frac{1}{8\pi}\int d^3r d^3r' J(\vec {r}){e^{-m_S|\vec{r}-\vec{r}\;'|} 
\over |\vec{r}-\vec{r}\;'|}J(\vec {r}\;').
\label{eq:e0r}
\end{equation}

A brief examination of Eq.~(\ref{eq:e0r}) shows that for 
values of $m_SR\gg1,$ $E_0$ is approximately
proportional to
$J_0^2 {4\pi\over 3} R^3$ times the volume integral of the 
Yukawa function, which means that, as expected, the binding energy is
proportional to the number of nucleons. This result would also be obtained 
if the theta function of Eq.~(\ref{eq:j}) were replaced by 
a smoother function that incorporated a nuclear surface thickness.
One sees also the usual result that the nucleus would not be stable
(the energy per nucleon would have no minimum as a function of $R$)
using scalar mesons alone.

One may proceed to evaluate $E_0$ by  using straightforward contour 
integration techniques with the result
\begin{eqnarray}
E_0& =& -{1\over 2}\left(g_S\rho_S\right)^2\left[
{4\pi\over 3} {R^3\over m_S^2}-2\pi{R^2\over m_S^3}\left(1-
{1\over m_S^2R^2}\right)\right]
\nonumber\\
& & -{1\over 2}\left(g_S\rho_S\right)^2
\left[-2\pi{R^2\over m_S^3} e^{-2m_SR}
\left(1+{1\over m_SR}\right)^2\right]
\nonumber\\
& &\stackrel{R \rightarrow \infty}{\longrightarrow}
-{1\over 2}\left(g_S\rho_S\right)^2\frac{4\pi}{3} \frac{R^3}{m_S^2}.
\label{energy}
\end{eqnarray}

The momentum distribution of scalar mesons is obtained by taking the 
necessary expectation value:
\begin{equation}
\rho ({\vec k}) \equiv \langle \psi_0|
a^\dagger_{\vec k} a_{\vec k} |\psi_0 \rangle
= \frac{ |\tilde{J}({\vec k})|^2 }{2\omega_{\vec k}^3}.
\end{equation}
Evaluation of Eq.~(\ref{eq:j}) using Eq.(\ref{jk}) gives 
\begin{eqnarray}
\tilde{J}({\vec k}) &=& \sqrt{\frac{2}{\pi}}J_0 \frac{
\sin (kR)-kR\cos (kR)}{k^3} \nonumber\\
&=& \sqrt{\frac{2}{\pi}}J_0R^3 {j_1(kR)\over kR}
\end{eqnarray}
where $k\equiv|{\vec k}|$, and $j_1(x)$ is the spherical Bessel
function of order one: $j_1(x)=\sin{(x)}/x^2-\cos{(x)}/x$. 
 Thus one finds
\begin{equation}
\rho ({\vec k}) = \frac{J_0^2}{\pi} R^6 \left({j_1(kR)\over kR}\right)^2
{1\over \omega_{\vec k}^3},
\label{rhorf}
\end{equation}
which is sharply peaked near $k=0$ for large nuclei, as
$j_1(x)$ has its first zero at $x=kR=4.5$.
One may compute the expectation value of the pion field in the state
$\mid\psi_0\rangle$ to find
\begin{eqnarray}
& \langle\psi_0\mid\phi(\vec{r})\mid\psi_0\rangle& \label{phir}\\
&=& \!\!\!\!\!\!\!\!\!\!\!\!\!\!\!
{g_S\rho_S\over 2m_S^2}
\left[2-{(1+m_S R)\over m_S r}e^{-m_SR}\left(e^{m_Sr}-e^{-m_Sr}\right)
\right],\nonumber 
\end{eqnarray}
for $r<R$. This function is a constant, ${g_S\rho_S\over 2 m_S^2}$
in the limit of infinite nuclear radius in which the conditions
$m_SR\gg1$ and $r\ll R$ hold.

\section{Meson Distribution in the LF Formulation-Scalar Mesons}
The essential observation for the LF formulation of systems
coupled to static sources is that static sources in a rest-frame
correspond to uniformly (constant $v^+=(v^0+v^3)/\sqrt{2}$) moving 
sources in a LF framework \cite{zako}.

The light front approach implies the following:
The canonical commutation relations
\begin{equation}
\left[ \phi(x^+\!,x^-\!,{\vec x}_\perp ),
\partial_-\phi(x^+\!,y^-\!,{\vec y}_\perp )\right]
=\frac{i}{2} \delta(x^-\!-y^-) \delta ({\vec x}_\perp-{\vec y}_\perp),
\label{can}
\end{equation}
where $x^\pm = (x^0\pm x^3)/\sqrt{2}$,
are satisfied by choosing
\begin{equation}
\phi(x^-\!,{\vec x}_\perp) = \int_0^\infty \!\!\!\!\frac{dk^+}{\sqrt{4\pi k^+}}
\int \!\frac{d^2k_\perp}{2\pi} \!\left[
a^\dagger_{k^+{\vec k}_\perp}e^{i{\bbox k}\cdot{\bbox x}}+
a_{k^+{\vec k}_\perp}e^{-i\bbox{k}\cdot\bbox{x}}\right] \label{lfexp}
\end{equation}
where 
$\bbox{k}\cdot\bbox{x}=k^+x^--\vec{k}_\perp\cdot\vec{x}_\perp$
as we are taking $x^+=0$. The operators 
$ a_{k^+{\vec k}_\perp}$ and $ a_{k^+{\vec k}_\perp}^\dagger$
obey 
standard boson commutation relations consistent with Eq.(\ref{can}).

A static charge distribution in a rest-frame
$J_{RF}({\vec x}_\perp,x^3)$ corresponds to a uniformly
moving charge distribution on the LF, where at $x^+=0$
\begin{equation}
J_{LF}({\vec x}_\perp,x^-)= J_{RF}({\vec x}_\perp,x^-v^+)=
J({\vec x}_\perp,x^3) .
\label{eq:jj}
\end{equation}
If the source is at rest in the rest-frame, then
$v^+=1/\sqrt{2}$ in Eq. (\ref{eq:jj}).
This relation between static external sources in a rest-frame and 
moving sources on the LF is exact. In order to see how it
arises, consider a time ($x^0$) independent scalar field
in LF variables [$x^\pm =(x^0\pm x^3)/\sqrt{2}$]
\begin{equation}
J_{LF}({\vec x}_\perp,x^-,x^+)
= J_{RF}({\vec x}_\perp,x^3=\frac{x^+-x^-}{\sqrt{2}},
x^0=\frac{x^++x^-}{\sqrt{2}})
\end{equation}
If the field is $x^0$-independent, then one finds for $x^+=0$:
$J_{LF}({\vec x}_\perp,x^-)= J_{RF}({\vec x}_\perp,-x^-/\sqrt{2})$,
which is Eq. (\ref{eq:jj}) for this parity-even  
spatial distribution and 
for the special case $v^+=1/\sqrt{2}$.

Using the fixed charge formalism from Refs. \cite{zako,mb:adv}
one thus finds for the LF Hamiltonian of scalar fields coupled
to this external source
\begin{eqnarray}
P^-&=&\sum_{k^+,{\vec k}_\perp}
\left[ \left(\frac{m_S^2+{\vec k}_\perp^2}{2k^+}
+\frac{k^+}{2{v^+}^2}\right) a_{k^+{\vec k}_\perp}^\dagger
a_{k^+{\vec k}_\perp}\right. 
\nonumber\\ 
& &+ \left.\frac{1}{v^+ \sqrt{2 k^+}}\left(\tilde{J}_{RF}(
{\vec k}_\perp,k^+/v^+)a_{k^+{\vec k}_\perp}^\dagger
+ h.c. \right)\right] . 
\label{eq:hlf}
\end{eqnarray} 
The term proportional to $\frac{k^+}{2{v^+}^2}$ arises from the plus momentum
of the heavy source and its inclusion is necessary to  maintain the
rotational invariance of the light-front approach; 
its derivation is presented in the  Appendix. 

The rest-frame binding energy can be obtained using
\cite{zako}
\begin{equation}
E_{RF} \equiv P^-v^+ .
\end{equation}
Like the rest frame Hamiltonian $H$ [Eq. (\ref{eq:h0})], 
the LF Hamiltonian $P^-$  [Eq. (\ref{eq:hlf})] 
is Gaussian in the scalar fields and the ground state is a coherent state
\begin{equation}
|\psi_0 \rangle_{LF} \propto \left[ 
\prod_{k^+,{\vec k}_\perp}
\exp \left( \frac{\tilde{J}_{RF}^*({\vec k}_\perp,k^+/v^+)
a_{k^+{\vec k}_\perp}^\dagger }
{v^+\sqrt{2k^+}\left(\frac{m^2+{\vec k}_\perp^2}{2k^+}+
\frac{k^+}{2{v^+}^2}\right) 
}\right)
\right]|0\rangle
\label{eq:psi0lf}
\end{equation}
with energy eigenvalue
\begin{eqnarray}
P^-_0= -\frac{1}{{v+}^2}\int_0^\infty \frac{dk^+}{2k^+}\int d^2k_\perp
\frac{|\tilde{J}_{RF}({\vec k}_\perp,k^+/v^+)|^2}
{\frac{m_S^2+{\vec k}_\perp^2}{2k^+}+\frac{k^+}{2{v^+}^2}} .
\label{eq:e0lf}
\end{eqnarray}
Note that the rest frame energy calculated from the LF result
Eq. (\ref{eq:e0lf}), i.e. $E_{RF}=v^+P^-_0$, and the binding
energy calculated directly in normal coordinates [Eq. (\ref{eq:e0})]
can be shown to be identical, as is necessary. To see this identity,
 make the change of integration variables:
${k^+\over v^+} \to k^3$. 

The LF-momentum distribution for the scalar mesons
\begin{equation}
\rho_S({\vec k}_\perp,k^+)
 =\langle\psi_0\mid 
a_{k^+{\vec k}_\perp}^\dagger
a_{k^+{\vec k}_\perp}\mid \psi_0\rangle
\end{equation}
can be calculated using 
Eq.~(\ref{eq:psi0lf}), with the result
\begin{equation}
\rho_S ({\vec k}_\perp,k^+)
= \frac{2k^+\left|\tilde{J}_{RF}({\vec k}_\perp,k^+/v^+)\right|^2}
{{v^+}^2\left[m_S^2+{\vec k}_\perp^2 +\left(\frac{k^+}{v^+}\right)^2
\right]^2 
} .
\label{eq:lfmom}
\end{equation}
Since $\tilde{J}_{RF}({\vec k}_\perp,k^+/v^+)$ is strongly
peaked for $k \sim 1/R$, the momentum distribution is also
trivially peaked near small momenta (for $R \gg 1/m_S$). 
Physically, this is
obvious since the mean field approximation gives rise
to meson fields which are nearly constant throughout the
nucleus. In Fourier space, a nearly constant meson field
corresponds to a momentum distribution which is peaked near
zero\cite{jerry}.

Note that the form of $\rho_S ({\vec k}_\perp,k^+)$ of Eq.~(\ref{eq:lfmom})
is different than the rest-frame momentum distribution $\rho(\vec{k})$
of  Eq.~(\ref{rhorf}). The light frame momentum distribution function and
the plus-momentum distribution obtained by integrating this quantity over
all values of $\vec{k}_\perp$ can only be obtained using the light front
formulation.

A quantity of great interest is the total LF momentum carried
by the meson field. A naive estimate of this quantity would yield 
momenta that are of the order of the (rest-frame)
binding energy times the velocity, i.e.
\begin{equation}
\langle k^+ \rangle_{naive} \approx v^+ |E_{RF}| .
\label{eq:kplusnaive}
\end{equation}
We will now show that this naive estimate is wrong in the mean field
approximation.

 Using Eq. (\ref{eq:lfmom}), one finds
\begin{eqnarray}
\langle k^+ \rangle &\equiv& \int d^2k_\perp \int_0^\infty dk^+ 
\rho ({\vec k}_\perp,k^+) k^+ \nonumber\\
&=& \int d^2k_\perp \int_0^\infty dk^+ 
\frac{2{k^+}^2\left|\tilde{J}_{RF}({\vec k}_\perp,k^+/v^+)\right|^2}
{{v^+}^2\left[m_S^2+{\vec k}_\perp^2 +\left(\frac{k^+}{v^+}\right)^2
\right]^2 } \nonumber\\ 
&=& v^+ \int d^3k 
\frac{k_z^2\left|\tilde{J}_{RF}({\vec k}_\perp,k_z)\right|^2}
{\left[m_S^2+{\vec k}^2 
\right]^2 } \nonumber\\
&=& \frac{v^+}{3}\int d^3k 
\frac{{\vec k}^2\left|\tilde{J}_{RF}({\vec k})\right|^2}
{\left[m_S^2+{\vec k}^2 
\right]^2 } ,
\label{eq:kplus}
\end{eqnarray}
where we used rotational invariance of the source in the rest-frame.
It is interesting to express the quantity $\langle k^+ \rangle $
in terms of a coordinate space integral analogous to Eq.~(\ref{eq:e0r})
for the energy. We find:
\begin{equation}
\langle k^+ \rangle  = 
\int\!\!\! d^3rd^3r' J(\vec {r})\!\!\left[\left(
1-{m_S\over2}|\vec{r}-\vec{r}\;'|\right){e^{-m_S|\vec{r}-\vec{r}\;'|}
\over 12\pi |\vec{r}-\vec{r}\;'|}\right]\!\! J(\vec {r}\;').
\label{eq:kpr}
\end{equation}
The quantity in brackets has a volume integral of 0 (with the
integration variable as $|\vec{r}-\vec{r}\;'|$). 
Thus the integral receives non-vanishing contributions only from 
regions near the nuclear surface, as is expected from the notion
that 
 the scalar meson field would be constant for a nucleus of infinite
 size
and so  receives nonzero Fourier components only from
the regions near the surface of the nucleus.
This means that $\langle k^+ \rangle  \propto R^2$ which gives a far
smaller
magnitude than the $R^3$ behavior of the binding energy. In
particular,   ${\langle k^+ \rangle \over v^+ E_0}\sim
{1\over R}$. This is in accord with Ref.~\cite{jerry} which shows that
$\langle k^+ \rangle$ vanishes for the case of infinite nuclear
matter.

We now make a more specific comparison
 of $\langle k^+ \rangle $ of Eq.~(\ref{eq:kplus})
with $v^+\mid E_{RF}|$ of Eqs.~(\ref{eq:e0}) and (\ref{eq:e0lf})
using
the specific model for the source $J({\vec r})$, Eq.~(\ref{eq:j}).
 A direct evaluation of the integral
yields 
\begin{eqnarray}
\langle k^+ \rangle&=&{v^+\over 3} \left(g_S\rho_S\right)^2\pi{R^4\over m_S}
\left[{1\over m_S^2R^2}-{3\over m_S^4R^4}\right.\nonumber\\
& &\left.+{e^{-2m_SR}\over m_SR} \left(1+{1\over m_SR}\right)
\left(1+{3\over 2 m_SR}+{3\over 2m_S^2R^2}\right)\right]\nonumber\\
& &\stackrel{R \rightarrow \infty}{\longrightarrow}
{v^+\over 3} \left(g_S\rho_S\right)^2\pi \frac{R^2}{m_S^3},
\end{eqnarray}
so that 
\begin{eqnarray}
{\langle k^+ \rangle\over v^+\mid E_0\mid}={1\over 2 m_SR}\approx{ 1\over 20}
\end{eqnarray}
in the limit of infinite $R$, and where the last estimate arises
from taking $m_S$ = 550 MeV. Thus the naive estimate fails.
For large values of $mR$ we see that the energy
grows like the volume of the system, while the plus momentum grows as
the area. 

The slow ${1\over 2 m_SR}$ approach to zero makes it interesting to study the
plus-momentum distribution of scalar mesons, which is the integral of 
$\rho_S(k_\perp,k^+)$ over all $\vec k_\perp$ divided by the nucleon number
$4\pi {R^3\over 3}\rho_B$ with $\rho_B$ the baryon density. 

Before considering
this quantity
it is worthwhile to follow the usual procedure of defining
a dimensionless momentum distribution function. This is done by using the
$x$ variable
\begin{equation}
x={k^+\over M_N v^+},\label{xdef}\end{equation}
which has the physical interpretation
of the LF-momentum distribution of mesons per nucleon, and taking the
nucleons plus-momentum to be $M_N v^+$, which occurs for a free nucleon at
rest. More precisely,
 one should use $M_N-15.75 $ MeV instead of $M_N$ to define the variable $x$
The latter number is
the average binding energy; practically speaking
this small shift  does not matter and is ignored here.

The variable $x$ is the quantity
of relevance for deep inelastic scattering from nuclear targets. We then have
after including the Jacobian of the transformation between the variables $k^+$
and $x$:
\begin{equation}
f_S(x)={3\over 4\pi R^3\rho_B}M_Nv^+\int d^2k_\perp\rho_S(k_\perp,k^+),
\end{equation}
which can be evaluated as (Fig. 1)
\begin{equation}
f_S(x)= M_N^2{\left(g_S\rho_S\right)^2 \over \rho_B}R^5{6\over \pi}
\;x\;
\int_{xM_NR}^\infty {dy\over \left(y^2+m_S^2R^2\right)^2} {j_1^2(y)\over y}.   
\label{result}
\end{equation}
\begin{figure}
\begin{Large}
\unitlength1.cm
\begin{picture}(15,6)(1.4,-7.2)
\includegraphics{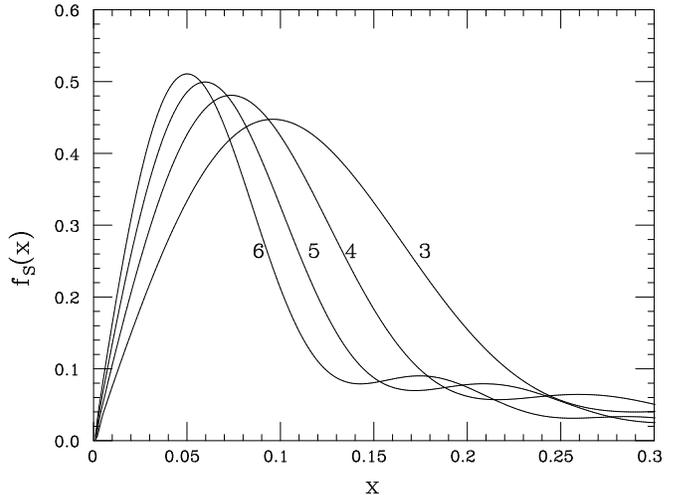}
\end{picture}
\end{Large}
\caption{The quantity $ f_S(x)$ vs. $x$ 
 for nuclei of radii 3,4,5,6   fm (the numbers
on the curves refer to the nuclear radius ). }
\label{fig:scalarf}
\end{figure}

It is interesting  to compare the result (\ref{result}) with previous
meson distributions obtained using the  equal time formulation of the
nuclear  wave function. These distributions exist  only for
pions\cite{et,jm}, but the  procedure outlined in Eqs.(22)-(24) of
Ref.\cite{jm} can be applied to scalar mesons.  Using the interaction
of Eq.~(\ref{eq:j}) and  the wave function of Eq.~(\ref{gsa}) in those
earlier formulae does indeed lead to the distribution function of 
Eq.~(\ref{result}). This is a consequence of the feature that the
interaction of Eq.~(\ref{eq:j}) does not excite the nucleus. Thus, in
this mean field example, the equal time ground state wave function is
sufficient to compute the distribution function. However,  the meson
distribution functions are light-like correlation functions which are
equal $x^-$ ``time" correlation functions, see e.g. the reviews
\cite{mb:adv,emcrevs}. Thus it is a specific and general feature of the light 
front wave approach that knowing only the ground state wave function
is  sufficient for computing  the distribution functions.

Armed with the distribution function $f_S(x)$ 
we may compute the momentum fraction carried by the nuclear scalar mesons
in our mean field approximation 
\begin{equation}
 \langle x_S\rangle \equiv \int_0^\infty x f_S(x) dx
\end{equation}
is not given by the
ratio of the binding energy per nucleon to the nucleon mass $M_N$, 
but by 
\begin{equation}
\langle x_S \rangle 
={\left(g_S\rho_S\right)^2 \over \rho_B m_S^3}\frac{3}{4R M_N},
\label{eq:x}
\end{equation}
in the limit of $m_SR\gg 1$. This quantity vanishes for nuclear matter. 

It is worthwhile to study how the limit of nuclear matter is approached.
The   numerical results to be presented are 
obtained using the parameters\cite{cw}
\begin{eqnarray}
\frac{g_S^2 M_N^2}{m_S^2} = 267.1,\qquad
 \rho_S= .179 fm^{-3}.
\end{eqnarray}
The  quantity $f_S(x)$ is displayed in Fig.(\ref{fig:scalarf}). One sees that 
$f_S(x)$ is fairly small for all $x$ and for all reasonable ranges of $R$.
The relevant size here is 2.5 which is the value of the nucleon distribution
function at its peak\cite{jerry} and which also is typical of 
peak values of the vector meson distribution discussed below.
Note that
the integral of $f_S(x)$  does not vanish. The quantity $xf_S(x)$ 
peaks at smaller and smaller values of $x$ 
for larger and larger values of $R$.
 
A final point concerns the expectation value of the scalar field. 
Defining the expectation value of the scalar field of the expansion
Eq.~(\ref{lfexp}) in the state $\mid\psi_0\rangle_{LF}$ as 
$\bar{\phi}_{LF}(x_\perp,x^-)$,
and computing the matrix element yields the result 
\begin{equation}
\bar{\phi}_{LF}(x_\perp,x^-)=
\langle\psi_0\mid\phi(r_\perp=x_\perp,r_3=-x^-v^+)\mid\psi_0\rangle,
\end{equation}
in which the latter matrix element is given in Eq.~(\ref{phir}).
Such a relation should be useful in constructing a more elaborate theory.
\section{Meson Distribution in the LF Formulation -- Vector Mesons}
The calculation of vector meson distributions is based on the
formalism in Ref. \cite{jerry}, which used earlier
work\cite{yansoper}.
 In the present work, we take the vector mesons
$V^\mu$ to be coupled to a large nuclear source of baryon current
$J^\mu$. Thus the relevant Lagrangian density ${\cal L}_V$ is given by
\begin{eqnarray}
{\cal L}_V =
-{1\over  4} V^{\mu\nu}V_{\mu\nu} +{m_V^2\over 2}V^\mu V_\mu 
-J^\mu \bar{V}_\mu\label{lag}
\end{eqnarray}
where the bare mass of the vector mesons given by
  $m_V$, and  $V^{\mu\nu}=
\partial ^\mu V^\nu-\partial^\nu V^\mu$ and\cite{jerry}
\begin{equation}
\bar{V}^\mu=V^\mu-\partial^\mu{1\over \partial^+}V^+
\end{equation} so that
$\bar{V}^+=0$.

The essential points which simplify our
analysis in the mean field approximation are
\begin{itemize}
\item The vector current is proportional to the 4-velocity of the
source, so that 
one finds
\begin{equation}
\tilde{J}^\mu_{LF} ({\vec k}_\perp, k^+) = v^\mu
\tilde{J}_{RF}^V({\vec k}_\perp, k^+/v^+),
\label{eq:jplus}
\end{equation}
where 
\begin{equation}
J_{RF}^V=g_V\rho_B \Theta(R-r),
\end{equation}
which has the same coordinate dependence as the model for $\tilde{J}_{RF}$
used  for scalar mesons and $g_V$ is the vector meson coupling constant.
\item We work in 
a frame where ${\vec v}_\perp=0$, and thus ${\vec J}_\perp=0$.
\item As explained in Ref. \cite{jerry}, we work with degrees 
of freedom such that the $+$ component of the vector meson field
vanishes.
\end{itemize}

This allows one to obtain the relevant Hamiltonian, which
can then be solved by 
means of a coherent state ansatz analogously to the case of 
scalar mesons. 

The main difference 
between vector mesons and scalar mesons is the appearance
of the polarization vector 
$\bar{\varepsilon}^\mu$ in the coupling of vector
mesons to the nucleon current \cite{jerry}
\begin{eqnarray}
\bar{V}^\mu = \int_0^\infty \frac{dk^+}{\sqrt{4\pi k^+}}
\int \frac{d^2k_\perp}{2\pi} \sum_{\omega=1,3} \bar{\varepsilon}^\mu 
({\vec k}, \omega)\times
\nonumber\\
\left[ a^\dagger_{k^+{\vec k}_\perp,\omega}e^{i{\bbox k}\cdot{\bbox x}}+
a_{k^+{\vec k}_\perp,\omega}e^{-i\bbox{k}\cdot\bbox{x}}\right]
\end{eqnarray}
where $ a_{k^+{\vec k}_\perp,\omega}$ satisfies
standard boson commutation relations 
and $\omega$ labels the polarization 
states.
Since the  transverse components of the nucleon current vanishes in
our model, and since $\bar{\varepsilon}^+=0$, the only relevant
component of $\bar{\varepsilon}^\mu$ is $\mu=-$. 

The coherent state  takes the form
\begin{equation}
|\psi_0^V \rangle \propto \left[ 
\prod_{k^+,{\vec k}_\perp}
\exp \left( \frac{\tilde{J}_{RF}^\mu({\vec k}_\perp,k^+/v^+)^*
\bar{\varepsilon}_\mu
a_{k^+{\vec k}_\perp}^\dagger,\omega}
{v^+\sqrt{2k^+}\left(\frac{m^2_V+{\vec k}_\perp^2}{2k^+}+
\frac{k^+}{2{v^+}^2}\right) 
}\right)
\right]|0\rangle.
\label{eq:psi0lfV}
\end{equation}
One may use this state to  take matrix elements of the plus-momentum operator
and the light front Hamiltonian. These expressions are similar to those for the
scalar mesons except that a factor of 
$(v^+\bar{\varepsilon}^-)^2 $ summed over all polarization
states is present. 
The polarization sum is given by \cite{jerry}
\begin{equation}
\sum_{\omega=1,3} \bar{\varepsilon}^-({\vec k},\omega)\bar{\varepsilon}^-
({\vec k},\omega) = \frac{m_V^2+{\vec k}_\perp^2}{{k^+}^2}.
\end{equation}

The light front Hamiltonian consists of three terms: the 
kinetic energy (Eq. (2.24) of Ref.~\cite{jerry}); the linear coupling and, the
effects of the instantaneous vector meson exchange. Thus  the light-front
energy of the 
nucleus due to coupling to the vector meson field consists of
two pieces. First there is a contribution which arises from physical vector
meson intermediate 
states, obtained from 
the expectation value of the kinetic energy and linear terms.
This piece 
differs from the meson momentum distribution by
one power of the 
energy denominator. In addition to this piece, the unphysical
components of the 
vector meson field contributes an instantaneous 
self-interaction of the nucleon field, which exactly cancels the most infrared
singular  terms of the term due to dynamical mesons. This instantaneous term is
very much analogous to the Coulomb interaction for an electromagnetic
field.
This term is obtained by canonical light front quantization and e.g.
is
included in Eq. (2.48) of Ref.~\cite{jerry}.

One takes the matrix element of the light front Hamiltonian to 
obtain the ground state energy of the vector meson field
coupled to the fixed source:
\begin{eqnarray}
P^-_0&=& P^-_{inst}-\frac{1}{{v^+}^2} 
\int_0^\infty\! \frac{dk^+}{2k^+} \!\int \!\!d^2k_\perp
\frac{ \frac{m_V^2+{\vec k}_\perp^2}{{k^+}^2}
\left| \tilde{J}_{LF}^+ ({\vec k}_\perp,k^+))
\right|^2}{\frac{m_V^2+{\vec k}_\perp^2}{2k^+}+
{1\over2}\frac{k^+}{{v^+}^2} }
\nonumber\\
&=& P^-_{inst}- \int_0^\infty \!\frac{dk^+}{2k^+} \!\int \!\!d^2k_\perp
\frac{\frac{m_V^2+{\vec k}_\perp^2}{{k^+}^2}
\left| \tilde{J}_{RF}^V({\vec k}_\perp,k^+/v^+))
\right|^2}{\frac{m_V^2+{\vec k}_\perp^2}{2k^+}+
{1\over2}\frac{k^+}{{v^+}^2}},
\nonumber\\ 
\label{eq:pmv}
\end{eqnarray}  
where we used Eq. (\ref{eq:jplus}) 
and where the instantaneous self-interaction is given by
\begin{equation}
P^-_{inst}= \int_0^\infty \frac{dk^+}{{k^+}^2} 
\int d^2k_\perp 
\left| \tilde{J}_{LF}^+ ({\vec k}_\perp,k^+))\right|^2 .
\label{eq:pminst}
\end{equation}
Note that the 
instantaneous piece [Eq. (\ref{eq:pminst}) can be combined with the rest 
in Eq. (\ref{eq:pmv}) yielding
\begin{equation}
P^-_0= \frac{1}{{v+}^2}\int_0^\infty \!\frac{dk^+}{2k^+}
\int \!\!d^2k_\perp
\frac{\left|\tilde{J}_{RF}^V({\vec k}_\perp,k^+/v^+)\right|^2}
{\frac{m_V^2+k_\perp^2}{2k^+}+{1\over2}\frac{k^+}{{v^+}^2}},
\label{eq:p0lf}
\end{equation}
which is  related to the vector meson contribution to the binding
energy $P^-_0v^+=E_{RF}^V$. The quantity 
$E_{RF}^V$, is up to a sign (reflecting the fact that
scalar meson give rise to attraction, while vector mesons give
rise to repulsion between nucleons) 
of the same form as the scalar result Eq. (\ref{eq:e0lf}).

The result of taking matrix elements of the plus-momentum operator in the 
coherent state leads to the result:
\begin{eqnarray}
\rho_V({\vec k}_\perp,k^+) &=& \frac{1}{2k^+}
\frac{ \frac{ m^2_V+{\vec k}_\perp^2}{{k^+}^2}
\left| \tilde{J}_{RF}^V ( {\vec k}_\perp,k^+/v^+)
\right|^2}
{\left[
\frac{m^2_V+{\vec k}_\perp^2}{2k^+}+
{1\over2}\left(\frac{k^+}{{v^+}^2}\right)^2\right]^2} 
\nonumber\\
&=& \left| \tilde{J}_{RF}^V \right|^2 \left\{
\frac{2}{k^+} \frac{1}{m^2_V+{\vec k}_\perp^2 + \left(\frac{k^+}{v^+}\right)^2}
\right.
\nonumber\\
& &\quad \quad \quad \quad \left.-
\frac{2k^+}{{v^+}^2} 
\frac{1}{\left[
m^2_V+{\vec k}_\perp^2 + \left(\frac{k^+}{v^+}\right)^2\right]^2}
\right\} .
\label{eq:rhov}
\end{eqnarray}

The second term on the r.h.s. of Eq. (\ref{eq:rhov}) is (up to a sign and a
differing mass) 
identical to the momentum distribution of scalar mesons. 
 The contribution of this second
term to the momentum (per nucleon)
carried by the vector mesons vanishes in the nuclear matter limit and 
explicit numerical evaluation shows that for finite nuclei
it is negligible compared with the
the first term.
This singular term
\begin{equation}
\rho_V^{sing}({\vec k}_\perp,k^+) \equiv
\frac{2\left| \tilde{J}_{RF}^V \right|^2 }{k^+} 
\frac{1}{m_V^2+{\vec k}_\perp^2 + \left(\frac{k^+}{v^+}\right)^2}\label{sing}
\end{equation}
is more interesting since it diverges as $k^+ \rightarrow 0$.
By direct comparison one can verify that the contribution
from this term to the total momentum carried by the vector mesons
is (up to a factor $v^+$) identical to the contribution 
of the vector mesons to the rest-frame energy of the nucleus
\begin{eqnarray}
\int_0^\infty \!\!\!dk^+ \!\!\!\int \!d^2k_\perp \rho_V^{sing} k^+
&=& \int_0^\infty \!\!\!dk^+ \!\!\int \!d^2k_\perp 
\frac{2\left| \tilde{J}_{RF} \right|^2  
}{m_V^2+{\vec k}_\perp^2 + \left(\frac{k^+}{v^+}\right)^2}
\nonumber\\
&=& {v^+}^2P^-_0 = v^+E_{RF}^V.
\end{eqnarray} 
Together with the result from the previous section that the non-singular
piece does not contribute to the momentum per nucleon in the nucleus
for infinite nuclear matter one thus finds that in the infinite nuclear
matter limit the momentum carried by the vector mesons is given by
\begin{equation}
\langle k^+_V \rangle = v^+ E_{RF},
\end{equation}
i.e. the momentum fraction carried by the vector mesons is given by
the ratio between the meson interaction energy and the mass of the
nucleus
\begin{equation}
\langle x_V \rangle = \frac{ \langle V \rangle}{M_A} .
\label{eq:xv}
\end{equation}
Note that the potential energy $\langle V \rangle$
on the right hand side of Eq. (\ref{eq:xv}) 
is only the part due to vector mesons. This result 
(momentum carried by vector mesons equals potential energy due to vector 
mesons) holds regardless whether or not there is also a scalar interaction
present. Since both scalar and vector interaction are rather large in
nuclei (of opposite sign, such that their net effect is small), this means
that vector mesons may carry a substantial fraction of the nucleus'
momentum. In the previous section, we have shown that scalar mesons
carry only a small fraction so the net momentum carried by the mesons
is essentially the momentum carried by the vector mesons, which can be
very large.

\subsection{Vector Meson Plus-Momentum Distribution}
The  result that vector mesons carry a large part of the nuclear
plus-momentum, which had also been predicted in Ref. \cite{jerry}
using more general arguments,
is at first surprising since the vector meson field (very much
like the scalar meson field) becomes space independent for nuclear
matter in the mean field approximation. Therefore, one would expect
that the vector meson field
for nuclear matter contains only quanta with vanishing $+$ momentum.
It thus seems paradoxical that vector mesons nevertheless
carry a finite fraction of the nucleus' $+$ momentum in this limit.

In order to resolve this apparent paradox, let us analyse the
momentum distribution arising from the crucial singular piece
$\rho_V^{sing}$ of Eq.~(\ref{sing}) more closely.

The dominant, 
singular part of the vector meson  light-front momentum density depends on
the source
\begin{eqnarray}
\tilde{J}_{RF}^V( {\vec k}_\perp,k^+/v^+)=
g_V\rho_B\sqrt{2\over\pi}R^3{j_1(kR)\over kR}
\end{eqnarray}
with 
\begin{equation}
k\equiv\sqrt{\vec {k}_\perp^2 +\frac{{k^+}^2}{{v^+}^2}}.
\end{equation}

We define a $+$-momentum distribution function per nucleon
$(A={4\pi\over 3}R^3\rho_B)$, using the $x$-variable of Eq.~(\ref{xdef})
 so that 
\begin{equation}
f_V(x)\equiv {3M_Nv^+\over 4\pi R^3\rho_B}
\int d^2k_\perp \rho_V^{sing}({\vec k}_\perp,k^+).
\end{equation}

One may simplify  the integral such that 
\begin{equation}
xf_V(x)=R^3{6\over \pi}{g_V^2\rho_B}
 \int_{xM_NR}^\infty {dy\over y}  {1\over y^2
+m^2_V R^2}j_1^2(y). \label{exact}
\end{equation}
A qualitative understanding may be gained by noting that the quantity 
$j^2_1(y)/y$ peaks at $y\approx1.6$, so that the denominator can be simplified
by treating ${1\over y^2+m_V^2R^2}\approx {1\over m_V^2R^2}$ and therefore 
\begin{equation}
xf_V(x)\approx{R\over m_V^2}{6\over \pi}  g_V^2\rho_B
 \int_{xM_NR}^\infty {dy\over y}j_1^2(y). \label{approx}
\end{equation}
The integral vanishes in the limit  $R\to\infty$ for all $x\ne0$.  In
particular, for $xM_NR\gg1$ an upper limit may be obtained by approximating
$j_1(y)\approx \cos{y}/y\ge 1/y$. Thus the integral is  less than
${1\over 2}({1\over xM_NR})^2$, so that for non-zero values of $x,$
\begin{equation} xf_V(x)\sim {R\over (xM_NR)^2},
\label{lim}
\end{equation}
which vanishes as $R$ approaches $\infty$.
If $x=0$,
the integral takes on the value 1/3. Thus in the limit of infinite  $R$
there is a sharp                     
distinction between the results for $x=0$ and for non-zero values,
no matter how small.
This suggests that $\lim_{R\to\infty} xf_V(x)$ might be a delta function.

To proceed further we need to show that 
\begin{equation}
\lim_{R\to\infty}\int_0^{x_U}dx xf_V(x) = ({g_V\over m_V})^2
{\rho_B\over M_N}
\label{non}
\end{equation}
and 
\begin{equation}
\lim_{R\to\infty}\int_{x_L}^{x_U}dx xf_V(x) = 0, \label{zero}
\end{equation}
for $x_L>0$. The above two equations would mean that
\begin{equation}
\lim_{R\to\infty} xf_V(x) = ({g_V\over m_V})^2{\rho_B\over M_N} \delta(x), 
\label{delta}
\end{equation}
in accord with the 
result expected from earlier work.
The present meaning of the function $\delta(x)$ is that integrals over $x$
including this delta function are non-vanishing provided the lower limit is
infinitesimally close to zero.
 
To verify Eq.(\ref{non}) it is simply necessary to do the integral.
One interchanges the order of integration,
takes $R\to\infty$ and uses the upper limit of the integral of
Eq.(\ref{approx}). Then the result is immediate if one uses the integral
$$
\int_0^\infty j_1^2(x) dx 
= \frac{\pi}{6}. 
$$
The verification of Eq.(\ref{zero}) also proceeds by integration and by 
using
Eqs.~(\ref{approx}) and (\ref{lim}).  Thus Eq.~(\ref{delta}) is valid    and
 in infinite nuclear matter the  quantity $xf_V(x)$ has
support only at $x=0$, or $k^+=0$.

The next step is to study the distribution function $f_V(x)$ of 
Eq.(\ref{exact}). First, we
calculate the momentum fraction carried by the vector mesons
\begin{eqnarray}
\langle x_V \rangle &\equiv& \int dx xf(x)
\nonumber\\
&=& \frac{1} 
{ M_N \frac{4 \pi}{3}R^3 \rho_B}
\int d^3k \frac{\left|\tilde{J}_{RF}({\vec k})\right|^2}
{m^2_V + {\vec k}^2}.
\end{eqnarray}
The integral is the same one that appears in the calculation
of the vector meson contribution ot the binding energy, showing that 
$\langle k^+\rangle=v^+E_{RF}$ for the vector meson contributions.
We take the limit of infinite nuclear matter to find
\begin{eqnarray}
\langle x_V \rangle  
&\stackrel{R \rightarrow \infty}{\longrightarrow}&
\frac{ 1}
{M_N 
\frac{4 \pi}{3}R^3 \rho_B}
\int d^3k \frac{\left|\tilde{J}_{RF}({\vec k})\right|^2}
{m_V^2}
\nonumber\\
&=& 
\frac{g_V^2 M_N^2}{m_V^2} \frac{\rho_B}{M_N^3 }
\end{eqnarray}
We follow Ref\cite{jerry} and use the parameters of 
of Chin and Walecka\cite{cw}
\begin{eqnarray}
\frac{g_V^2 M_N^2}{m_V^2} = 195.9,\qquad
 \rho_B= .193 fm^{-3}.
\end{eqnarray}
Then $\langle x_V \rangle =0.348,$
which agrees with the result of Ref\cite{jerry}.

\begin{figure}
\begin{Large}
\unitlength1.cm
\begin{picture}(15,6.3)(1.4,-7.2)
\includegraphics{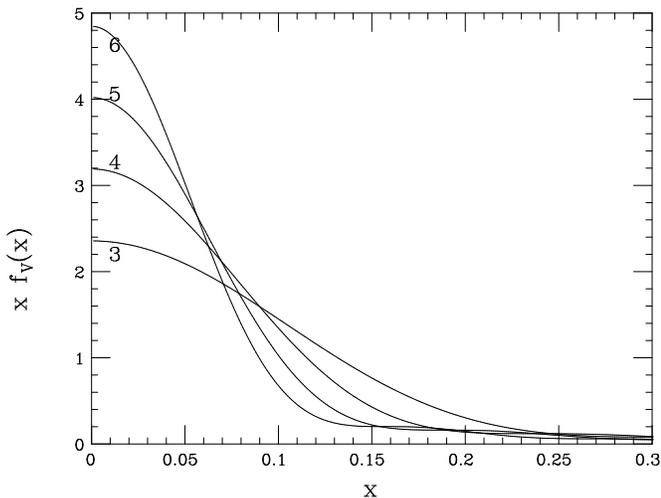}
\end{picture}
\end{Large}
\caption{The quantity $x f_V(x)$ vs. $x$ 
 for nuclei of radii 3,4,5,6   fm (the numbers
on the curves refer to the nuclear radius ). }
\label{fig:vectorf}
\end{figure}

Numerical studies of $xf_V(x)$ of 
      Eq.~(\ref{exact}) for finite values of  $R$ 
are shown in Fig.~\ref{fig:vectorf}.

The results of Fig.~\ref{fig:vectorf}
 show the typical behavior of distributions that approach 
delta functions.  The approach to the delta function limit is slow, as
expected
from the $1/R$ behavior displayed in Eq.~(\ref{lim}). 
Indeed, all of the
results show a significant spread and the functions do have some support for 
values of $x$ such that $x$ greater than about 0.1.
The key physics question we need to address is whether or not the 
distribution is non-zero for values of $x$ that are too small to be
observable.  This does happen in the limit of infinite nuclear matter 
$R\to\infty$.  We may examine this question in another way by considering
 how much of the total vector meson plus momentum occurs in an observable
 region.
We arbitrarily define the low $x$ (un-observable) region here  as $x<0.1$.
The distribution functions shown here are to be convoluted with a vector-meson
quark-distribution function, which peaks (qualitatively) at 1/2. Thus 
$x<0.1$ corresponds to $x_{Bj}$ less than about 0.05. That value is small
enough so that we can say the vector mesons would be hidden by the effects of 
shadowing. Even in that case we see qualitatively 
from Fig.~(\ref{fig:vectorf}) and by computation that approximately
0.05 (R=6 fm) and 0.08 (R=3 fm)  of
the total nuclear plus momentum occurs in the observable region of 
$x_{Bj}>0.05$. Thus the mean field vector
mesons could be observable in this model. We note that the 
parameters used here 
have been shown  to give a much larger nucleon depletion 
than allowed by data. Thus the current results provide an overestimate.

\section{summary}
We have explicitly constructed the scalar and vector meson states
 for a mean field
model of large nuclei  and have obtained some rigorous results for the 
meson distribution functions.

Direct calculation of the light-front
momentum distribution for scalar mesons in the mean field approximation
is localized near $k^+\rightarrow 0$. As a result, the total
momentum carried by the scalar mesons in the mean field approximation
is vanishingly small for nuclear matter.
This calculation thus confirms the results of Ref. \cite{jerry}:
even though scalar mesons contribute to the nuclear binding in nuclear
matter, they carry only a vanishing fraction of the momentum in
the mean field calculation. 

Also in accord with Ref.~\cite{jerry} is the present result that
vector mesons do contribute to the plus-momentum of
the nucleus in the same limit of infinite nuclear matter. 
In fact, we show that
the momentum fraction carried by vector mesons in mean field approximation
and in the nuclear matter limit is given by the ratio between the 
the vector meson contribution  to the potential energy and
the nuclear mass.
Surprisingly, this result is obtained despite the fact that the
vector meson distribution functions are zero for
non-zero $x_{Bj}$ for infinitely large nuclei. 
This is because the vector meson distribution function is sharply peaked
such that $xf_V(x)\propto \delta(x)$.

For nuclear radii $R$ corresponding to realistic large nuclei,
the meson distributions, while strongly peaked at low values of $x$,
are wide enough so that some fraction of the mesons cold be observable.
Thus the strict zeros, of Ref.~\cite{jerry}, are not obtained. However, the
predictive power of the momentum sum rule is  still vitiated because
a significant fraction of the mesons are hidden at small values of $x_{Bj}$.

\appendix
\section{Derivation of the Recoil Light Front Hamiltonian}
This Appendix is concerned with the derivation of Eq.~(\ref{eq:hlf}).
This is implicit in the work of Ref.~\cite{zako} and also in 
Ref.~\cite{gh}, but the importance of this term makes it relevant to derive it
for the current nuclear context.

Consider a situation in which a nucleus $\mid\Psi_A\rangle$ consists of
one heavy source of mass $M_H$ and light degrees of freedom $L$. 
The light front momentum and energy eigenvalues are $P_A^\pm=M_Av^\pm$ with
$v^+v^-={1\over2}$. It is worthwhile to remove the large mass $M_H$ from the
eigenvalue. Thus we follow Ref. \cite{zako} and write
\begin{equation}
M_A=M_H+\delta E,
\end{equation}
where $\delta E$ is the binding energy.
The light
front eigenvalue equation $M_A v^-\mid\Psi_A\rangle=
\widehat P^-\mid\Psi_A\rangle$
then reads
\begin{eqnarray}
&&{M_H+\delta E\over 2v^+}\mid\Psi_A\rangle \label{eig} \\
&&\quad \quad \quad  =
\int d^2k_\perp dk^+{k^2_\perp+M_H^2\over 2 k^+}
A^\dagger(k_\perp,k^+)A(k_\perp,k^+)\mid\Psi_A\rangle \nonumber\\
&&\quad \quad \quad\quad +(\widehat P^-_{HL} +\widehat P^-_{LL})\mid\Psi_A\rangle \nonumber
\end{eqnarray}
where $A(k_\perp,k^+)$ is the destruction operator of the heavy  source,
and the light front Hamiltonian for the interaction between the heavy and
light degrees of freedom is $\widehat P^-_{HL} $ 
and that for the light degrees of
freedom is $\widehat P^-_{LL} $. 
One would like to remove the large mass $M_H$ from
both sides of the equation. This may be accomplished by 
making the replacement 
\begin{eqnarray}
k^+\to P_A^+ -  \widehat P^+_L \nonumber\
=M_Hv^+\left(1+{\delta Ev^+-\widehat P^+_L\over M_Hv^+}\right)
\end{eqnarray}
in Eq.~(\ref{eig}), where $\widehat P^+_L$ 
is the second-quantized kinematic-plus-momentum of the light degrees of
freedom. Using the above replacement, and ignoring terms of order $M_H^{-1}$
leads to the replacement
\begin{equation}
{k^2_\perp+M_H^2\over  k^+}\to{M_H\over v^+}-{\delta E\over v^+} +
{\widehat P^+_L\over {v^+}^2}
\end{equation} The next step is to remove the heavy source of momentum 
$k_\perp,k^+$ from $\mid\Psi_A\rangle$ by acting on it with the operator
$A(k_\perp,k^+)$. The result is that Eq.~(\ref{eig})  is transformed into
\begin{eqnarray}
&&{2\delta E\over v^+}A(k_\perp,k^+)\mid\Psi_A\rangle\\
&&\quad \quad \quad \quad \quad =
\left({\widehat P_L^+\over {v^+}^2} +
2\widehat P^-_{HL} +2\widehat P^-_{LL}\right)
A(k_\perp,k^+)\mid\Psi_A\rangle. \nonumber
\end{eqnarray}
The operator ${\widehat P_L^+\over 2{v^+}^2}$ which is the contribution of the
recoil momentum of the heavy source to the light front Hamiltonian is the
operator that we seek. Note that the quantity in parenthesis and 
${2\delta E\over v^+}$ are independent of $k_\perp, k^+$.

\end{document}